# Beauty and the Beast:
# What Lattice QCD Can Do for $B$ Physics

Andreas S. KRONFELD

*Theoretical Physics Group, Fermi National Accelerator Laboratory,
P.O. Box 500, Batavia, IL 60510, USA*

18 September 1993

ABSTRACT

The role lattice QCD can play in $B$ physics is surveyed. We include results for the decay constant, and discuss upcoming calculations of semileptonic form factors and neutral-meson mixing. Together with experimental measurements, these calculations can determine the unitarity triangle.

hep-ph/9310220  18 Oct 93

# 1 INTRODUCTION

One of the reasons why $b$-hadrons are interesting is that their properties (decays, mixing, $CP$ violation) help determine the least well-known elements of the Cabibbo-Kobayashi-Maskawa (CKM) matrix. (For a review of the CKM matrix, see Ref. 1.) Leptonic and semileptonic $B$-meson decay amplitudes are proportional to the CKM matrix elements $V_{cb}$ or $V_{ub}$. Through top-quark box diagrams, $B_q^0$-$\bar{B}_q^0$ mixing is sensitive to $V_{tq}$, where $q$ denotes a $d$ or an $s$ quark. In each case, however, the standard-model expression for the (differential) decay rate follows the pattern

$$\begin{pmatrix} \text{experimental} \\ \text{measurement} \end{pmatrix} = \begin{bmatrix} \text{known} \\ \text{factors} \end{bmatrix} \begin{pmatrix} \text{QCD} \\ \text{factor} \end{pmatrix} \begin{pmatrix} \text{CKM} \\ \text{factor} \end{pmatrix} \qquad (1.1)$$

The known factors consist of well-known constants and experimentally measurable quantities such as masses and kinematic variables. But, as a rule, the QCD factor is nonperturbative and cannot be deduced from other experiments. Therefore, to extract the CKM factor from the measurement one must have reliable theoretical calculations in nonperturbative QCD.

The only systematic, first-principles approach to nonperturbative QCD is the formulation on the lattice.[2] The most promising calculational method has proven to be large-scale numerical computations. Much like an experimentalist, a lattice theorist must contend with statistical and systematic errors in numerical data. Hence, the reliability of the calculation boils down to the care and control of the uncertainties. Only recently, however, have methods and machines become powerful enough to produce reasonably reliable estimates for the quantities needed to pin down standard-model parameters. Although this report focuses on $B$ physics, a recent review is more general.[3]

How does lattice QCD compare to other theoretical approaches to properties of $b$ hadrons? The main strength of lattice QCD is that it *is* QCD. Given enough computing resources the numerical results are derived from the first principles of the path integral, the renormalization group and the QCD Lagrangian. There are only $n_f + 1$ free parameters, corresponding to quark masses and the gauge coupling. Once these are fixed by experiment, using meson masses to fix the quark masses and the 1P–1S splitting of quarkonium* $\Delta m_{\text{1P–1S}}$ to fix $\Lambda_{\text{QCD}}$, there are no more adjustable parameters. By contrast, both QCD sum rules and effective field theories introduce additional parameters—condensates or coupling constants, respectively, which are not calculable in a self-consistent fashion.

Of course, numerical lattice QCD is not omnipotent. Computational physics is more labor intensive than theoretical physics, though less so than experimental physics. In the case of lattice QCD, the field is just starting to mature. Other aspects of the numerical technique—imaginary time and the finite volume—make some calculations less feasible. Nevertheless, the origins of the uncertainties in the numerical calculations are conceptually understood. In $B$ physics results for leptonic and semileptonic decays and neutral-meson mixing are limited only by computer and human resources. But

---
*This quantity is especially *in*sensitive to the quark masses.

by the end of the decade the uncertainty in the QCD factor of eq. (1.1) for these measurments should be less than or comparable to the experimental uncertainties.

For the time being, one must live with something called the "quenched approximation" (cf. sect. 2), if other errors are to be brought under control. The quenched approximation is easy to describe: it omits the vacuum polarization of the quarks. For heavy quarks ($c$, $b$, $t$) this is probably tolerable, because their vacuum polarization is short-distance, and hence mostly perturbative. Similarly, one ought to be able to compensate for short-distance, light-quark ($u$, $d$, $s$) vacuum polarization. Long-distance effects of light quarks is harder to characterize. Nevertheless, the quenched approximation can be hoped to provide a useful phenomenology, because it embodies more of QCD than, say, the naive quark models do. But, as with an empirical model, presuming predictions in one arena after success in another may be subject to trial and error.

This paper is organized as follows: Because of the importance of the uncertainty estimates sect. 2 reviews some of the theoretical foundation and the origin of systematic errors in the numerical calculations. To illustrate the advantages of a systematic approach, recent calculations of light hadron masses[4] and decay constants[5] are briefly discussed in sect. 3. The emphasis of sect. 4 is on properties of the $B$ meson—leptonic (sect. 4.1) and semileptonic (sect. 4.2) decays and neutral-meson mixing (sect. 4.3)—for which reliable QCD calculations will be available within the next few years. Prospects for studying nonleptonic decays are discussed in sect. 4.4, and results on the $B$ meson wave function are mentioned in sect. 4.5. Sect. 5 shows how a combination of experimental measurements and the lattice QCD calculations discussed in sect. 4 can be assembled to determine the sides of the celebrated unitarity triangle. Together with the assumption of 3-generation unitarity (i.e., the unitarity polygon is indeed a triangle), the three sides yield the angles $\alpha$, $\beta$, and $\gamma$ describing $CP$ asymmetries.

## 2 THEORETICAL AND NUMERICAL BASICS

According to Feynman, vacuum expectation values can be represented as a path integral. In field theory, a mathematically sound definition starts with a lattice of finite volume, depicted in Fig. 1. For QCD the degrees of freedom are gluons $A_\mu^a(x)$ ($a$ is a color index), quarks $\psi_i(x)$ ($i$ is an index for spin, color, and flavor), and anti-quarks $\bar\psi_i(x)$. Then an expectation value is given by

$$\langle \bullet \rangle = \lim_{L\to\infty} \lim_{\substack{a \to 0 \\ L = Na \text{ fixed}}} \frac{1}{Z_{L,a}} \int \prod_{x,\mu,a} dA_\mu^a(x) \prod_{x,i} d\psi_i(x) \prod_{x,i} d\bar\psi_i(x) \bullet e^{-S(A,\psi,\bar\psi)}, \quad (2.1)$$

where $S$ is (a lattice version of) the QCD action. The normalization factor $Z_{L,a}$ is defined so that $\langle 1 \rangle = 1$ for each $L$ and $a$.

As an application of eq. (2.1), let $\hat\Phi$ denote an operator with well-specified quantum numbers, built out of $A_\mu^a$, $\psi_i$, and $\bar\psi_i$, and consider

$$\langle \Phi(t)\Phi^\dagger(0) \rangle = \langle 0|\hat\Phi e^{-\hat H t}\hat\Phi^\dagger|0\rangle. \quad (2.2)$$

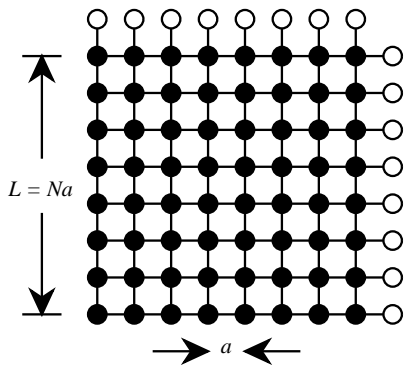

Figure 1: The finite lattice consists of a discrete set of points $x$ separated by lattice spacing $a$. If the number of points on each side is $N$, the linear size of the finite volume is $L = Na$. Usually one uses periodic boundary conditions, which would identify the white sites at the top (far right) with the sites at the bottom (far left).

Note that the evolution is through imaginary time ($e^{-\hat{H}t}$ instead of $e^{i\hat{H}t}$), which makes the integral in eq. (2.1) converge better (weight $e^{-S}$ instead of $e^{iS}$). Inserting complete sets of states

$$\langle \Phi(t)\Phi^{\dagger}(0)\rangle = \sum_n \left|\langle 0|\hat{\Phi}|n\rangle\right|^2 e^{-E_n t} \stackrel{\text{large } t}{\approx} \left|\langle 0|\hat{\Phi}|1\rangle\right|^2 e^{-E_1 t}, \qquad (2.3)$$

where $E_n$ is the energy of the $n$-th state. For large enough $t$ the lowest-lying state dominates, so its energy $E_1$ can be read off from the exponential fall-off. If $\Phi$ has the quantum numbers of a $B$ meson at rest, then $E_1 = m_B$. By a similar approach, one can determine matrix elements. Substituting a current $J$ for $\Phi$ in eq. (2.2) yields

$$\langle J(t)\Phi^{\dagger}(0)\rangle \stackrel{\text{large } t}{\approx} \langle 0|\hat{J}|1\rangle\langle 1|\hat{\Phi}^{\dagger}|0\rangle e^{-E_1 t}, \qquad (2.4)$$

for large $t$. Once $E_1$ and $\langle 1|\hat{\Phi}^{\dagger}|0\rangle$ have been determined from eq. (2.3), eq. (2.4) yields $\langle 0|\hat{J}|1\rangle$. If $J$ is the charged weak current and $\Phi$ again has the quantum numbers of a $B$ meson at rest, $\langle 0|\hat{J}|1\rangle$ is proportional to $f_B$, cf. sect. 4.1. In an obvious jargon, eqs. (2.2) and (2.4) are called two-point functions. For matrix elements with hadrons in the final state too, one calculates a *three*-point function

$$\langle \Phi_f(t_1)J(t_2)\Phi_i^{\dagger}(0)\rangle \stackrel{\text{large } t_1,t_2}{\approx} \langle 0|\hat{\Phi}_f|f_1\rangle\langle f_1|\hat{J}|i_1\rangle\langle i_1|\hat{\Phi}_i^{\dagger}|0\rangle e^{-E_{f_1}t_1 - E_{i_1}t_2}, \qquad (2.5)$$

to obtain $\langle f_1|\hat{J}|i_1\rangle$. Matrix element of this kind are needed for semileptonic form factors and neutral-meson mixing.

Nonperturbative calculations of eqs. (2.2), (2.4), and (2.5) actually yield masses and matrix elements in "lattice units," e.g. $am_B$ rather than $m_B$. Physical results are obtained by extrapolating dimensionless ratios. For example,

$$\frac{f_B}{m_B} = \lim_{L\to\infty} \lim_{\substack{a \to 0 \\ L = Na \text{ fixed}}} \frac{af_B(L,a)}{am_B(L,a)}. \qquad (2.6)$$

Fortunately, both limits are constrained by theoretical considerations. The infinite-volume limit $L \to \infty$ must conform with general properties of massive quantum field theories in a finite volume.[6] In QCD the *pattern* of approach to the continuum limit $\genfrac{}{}{0pt}{}{a \to 0}{L = Na \text{ fixed}}$ can be deduced from perturbation theory, because of asymptotic freedom.

Familiar units of MeV are restored by using a standard mass in the denominator of eq. (2.6) and setting it to its physical value. Owing to the renormalization group, this equivalent to eliminating the bare gauge coupling, one of the free parameters of QCD. Rather than $m_B$, as indicated in eq. (2.6), typical choices are $m_\rho$ or the 1P–1S splitting of quarkonium $\Delta m_{1P-1S}$. The latter is especially insensitive to the quark masses, i.e. the other parameters of QCD. The quark masses are also parameters that must be set by experimental input. For example, $m_b$ is fixed by tuning $m_\Upsilon/\Delta m_{1P-1S}$ to its physical value.

Eq. (2.1) makes an explicit mathematical analogy between quantum field theory and statistical mechanics. Starting from eq. (2.1), therefore, a wide variety of nonperturbative techniques from statistical physics can be applied to field theory. For QCD the most promising has proven to be a numerical method. First $a$ and $L$ are fixed. Then the left-hand-sides of eqs. (2.2), (2.4), and (2.5) are merely integrals of a finite, though huge, dimension $\sim (L/a)^4 \times 4 \times 8$. In practice, available memory in the largest supercomputers limits the dimension to $10^7$–$10^{10}$. The only practical way to evaluate integrals of such high dimension is Monte Carlo integration with importance sampling, almost always with weight $e^{-S}$. Then the whole procedure is repeated for a sequence of $a$'s holding $L = Na$ fixed, and for sequences of $L$'s holding $a$ fixed.

There are two ways to reduce the statistical errors. One is to carry out longer Monte Carlo runs. This puts a premium on computer speed. The other is to choose the largely arbitrary operator $\Phi$, above, to maximize the signal-to-noise ratio of the two- and three-point functions. This puts a premium on computer programmability. From eq. (2.6) it is clear that the statistical errors must be under control if sensible extrapolations in $a$ and $L$ are to be made.

There are also two ways to take the continuum limit, and, hence, to control finite lattice-spacing errors. One is by brute force, making $a$ smaller and smaller, using a simple form of the action $S$. The other way, which should save computer time, is to improve the accuracy of the lattice action. This is the generalization to field theory of methods familiar from the numerical solution of differential equations. In the past, statistical errors were often too large to notice any practical improvement from this theoretical improvement. Now, however, there are several examples, and one should expect "improved actions" to play an important role in $B$ physics.

For complicated technical reasons the most time-consuming part of the numerical calculations involve treating the light quarks. The physical root of these problems is the Pauli principle: a fermion over here always "knows" something about a fermion way over there. It turns out that one can save a factor of $10^2$–$10^3$ in computer time by neglecting the back reaction of quarks on the gluons. As mentioned in the Introduction, this amounts to omitting vacuum polarization while treating the interaction between the valence quarks and the gluons exactly. This approximation is therefore sometimes called the valence approximation. More often it is called the quenched approximation (calling on an technical analogy to condensed matter physics). Fig. 2 illustrates examples of quark-flow diagrams that are kept (a, c) or omitted (b, d) in the quenched approximation. In particular, the quenched approximation spoils the mechanism generating the mass of the $\eta'$, with consequences that could affect other masses through self-energy interactions.[7]

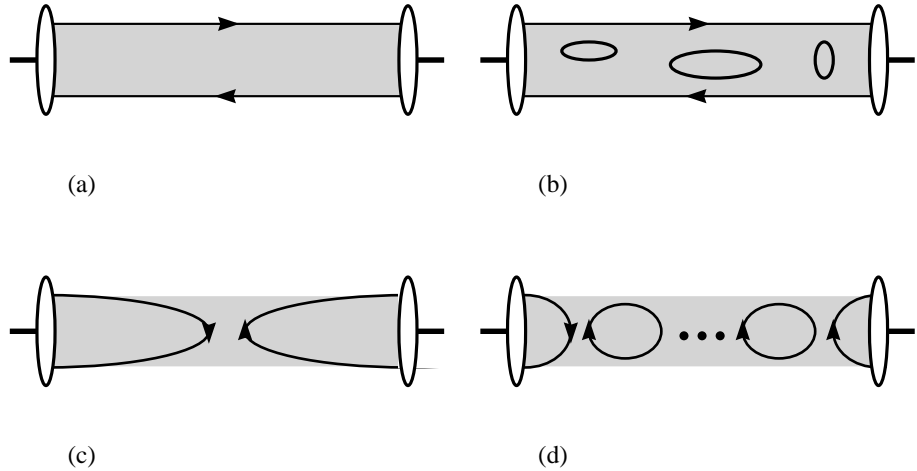

(a)  (b)

(c)  (d)

Figure 2: (a) A meson consisting of valence quarks (lines) interacting with the glue (gray shading); this quark-line topology is kept in the quenched approximation. (b) Same as (a) but with some sea quarks; this topology is omitted in the quenched approximation. (c) A flavor-singlet topology kept in the quenched approximation. (d) Flavor-singlet topologies omitted in the quenched approximation; such diagrams generate the $\eta'$ mass.

Another way to assess the quenched approximation is at the quark-gluon level. As shown in Fig. 3, the gauge coupling runs too quickly in the quenched approximation. In quenched QCD one effectively adjusts the quenched gauge coupling (dotted line in Fig. 3) at the cutoff, so that it agrees with the real coupling (solid line in Fig. 3) at the scale of the physics (denoted $\mu_{ph}$ in Fig. 3). If the quenched approximation is at all successful, many quantities with typical scale $\mu_{ph}$ should be verifiable. On the other hand, one need not expect quantities with a typical scale rather different from $\mu_{ph}$ to be verified. Usually this consideration is merely heuristic. For nonrelativistic systems, i.e. the $\psi$ and $\Upsilon$ families, the two-body wave function provides the probability of each scale, so one can account quantitatively for the effects of the difference between the two curves.[8]

To conclude this section, let us offer a handful of questions the nonexpert should keep in mind when appraising lattice QCD calculations:

1. Are the statistical errors small enough to understand anything?

2. Is the lattice spacing large enough? Or, even better, have lattice-spacing errors been extrapolated away?

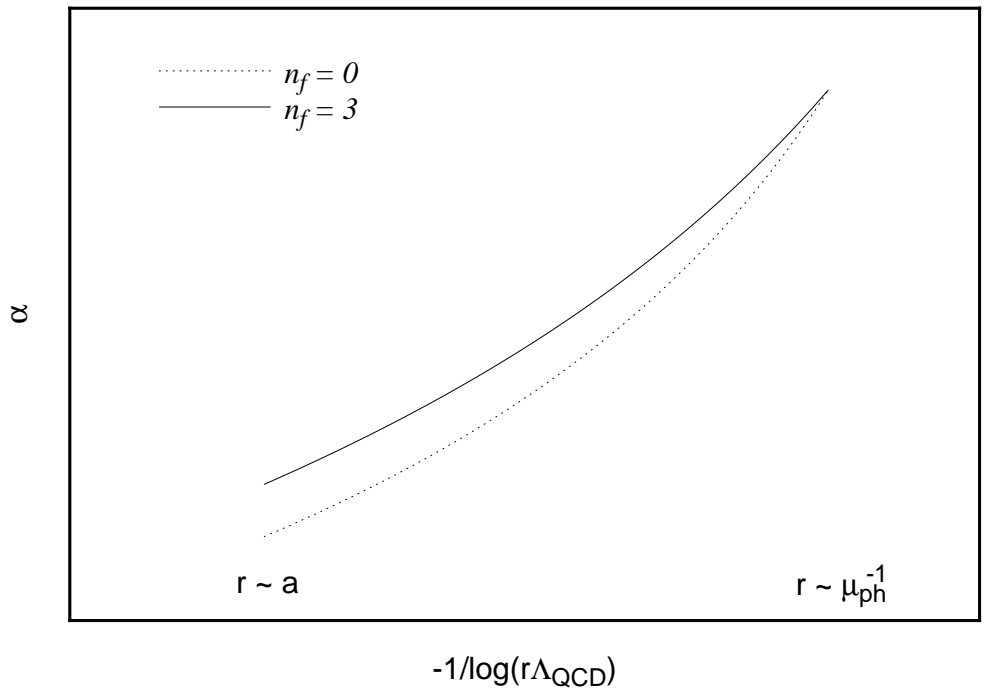

Figure 3: Sketch of the gauge coupling in quenched and "full" QCD. The flavor-dependence of the $\beta$-function coefficient $b_0 = 11 - 2n_f/3$ implies that the coupling in the quenched ($n_f = 0$) case runs more quickly at short distances.

3. Is the physical volume large enough? Or, even better, have finite-volume errors been extrapolated away?
4. Have the quark masses been adjusted precisely enough?
5. Is the quenched approximation acceptable?

With a little luck the lattice mavens will always answer, "At the $x\%$ level, yes."

## 3 LIGHT HADRON SPECTRUM

One of the original goals of lattice QCD was a first principles calculation of the light hadron mass spectrum. A recent paper[4] employing the quenched approximation reports a significant step towards that goal. Using the GF-11, a special purpose computer designed at IBM,[9] Weingarten, et al, have evaluated the path integral at three values of $a$ (and fixed $L$) and, at the coarsest lattice spacing, three values of $L$. With $m_\rho$ to convert from lattice units to MeV (cf. eq. (2.6)) and $m_\pi$ and $m_K$ to set the light and strange quark masses, their results for two vector mesons and six baryons are summarized in Fig. 4. The error bars represent the authors' estimates of the accumulated uncertainties

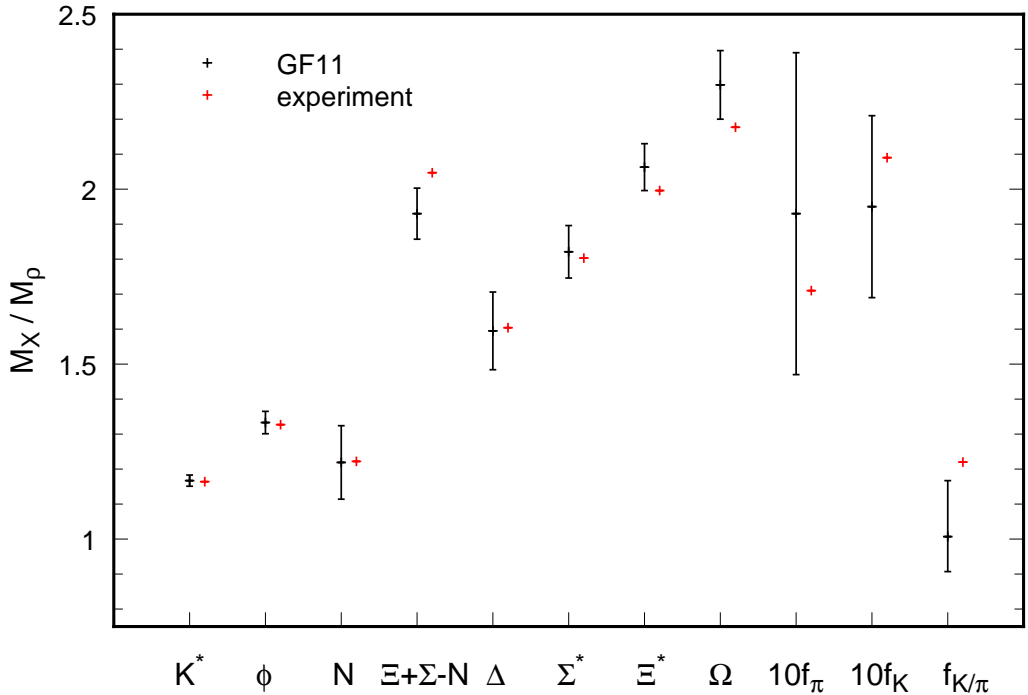

Figure 4: Spectrum and decay constants of the light hadrons. Error bars are from lattice calculations in the quenched approximation,[4, 5] and + denotes experiment.

from all sources except the quenched approximation.

The agreement between these quenched QCD results and nature is tantalizing. Experts[3, 10, 11] in the field might quibble about some details of the analysis, but they cannot deny that such a systematic attack on the errors is basically sound. A "bottom-line" example is the ratio $m_N/m_\rho$, which without extrapolation is too large.[10] After extrapolation, however, this ratio agrees to an accuracy much better than the quoted precision. Moreover, there are some nontrivial cross-checks: The value of $\Lambda_{\text{lat}}^{(0)}$ ($=\Lambda_{\text{QCD}}$ in the "lattice" scheme with $n_f = 0$ active flavors) agrees with the value obtained in lattice QCD studies of charmonium.[8, 12] In charmonium, however, it is possible to correct for the quenched approximation, because most of the error comes from short distances. The same calculations[8] obtain a value of $\Lambda_{\overline{\text{MS}}}^{(4)}$ that agrees with deep, inelastic scattering and other high-energy processes.

Fig. 4 also shows results for the $\pi$ and $K$ decay constants.[5] We have converted the results to the convention of eq. (4.2), below, in which $f_\pi = 131$ MeV. The relative uncertainties are larger than for the mass ratios. Because the decay constants are more sensitive to short distances, one might hope that the ratio $f_K/f_\pi$ would be less sensitive to the errors of the quenched approximation. Unfortunately, the numerical results do not support this idea.

# 4  *B*-PHYSICS

In contrast to the light hadron physics discussed above, the lattice-spacing and finite-volume dependence of $B$ meson properties has not yet been thoroughly investigated. An exception to this rule is the study of the decay constant in the theoretically interesting limit of an infinitely heavy $b$ quark.[13, 14] This limit is often called the static limit, because the heavy quark is anchored in one place. It seems, however, that the $1/m_b$ correction to the decay constant is large, so that these results are not directly applicable to phenomenology.

The dynamics of a hadron with one heavy quark is surprisingly simple, because the energy scale associated with the heavy quark mass decouples. For this reason, it is possible to treat a heavy quark on the lattice,[15, 16, 17] even when $m_q a \sim 1$.

In this section, subsection titles indicate the product of CKM matrix element and $B$-meson property, where appropriate.

## 4.1  Leptonic Decays: $f_B|V_{ub}|$

The leptonic width of the charged $B$ meson is given by

$$\Gamma[B \to l\nu] = \left[\frac{G_F^2 m_l^2 m_B}{8\pi}\eta_{\rm em}\left(1 - \frac{m_l^2}{m_B^2}\right)\right] f_B^2 |V_{ub}|^2. \qquad (4.1)$$

Eq. (4.1) is a concrete example of eq. (1.1). The numerical value of the bracket is well known, although the electromagnetic radiative correction $\eta_{\rm em}$ is uncertain at the 0.1% level. To determine $|V_{ub}|$ through a measurement[*] of a leptonic decay, one must first know the decay constant $f_B$, defined by

$$\langle 0|\bar{u}\gamma_\mu\gamma_5 b|B^-(p)\rangle = ip_\mu f_B \qquad (4.2)$$

with the normalization convention $\langle B^-(q)|B^-(p)\rangle = 2E_B(2\pi)^3\delta^{(3)}(\bm{p} - \bm{q})$.

The two-point function in eq. (4.2) [cf. eq. (2.4)] is one of the most straightforward of lattice QCD calculations. A recent preprint,[18] for example, finds

$$\begin{aligned}
f_B &= 187(10) \pm 12 \pm 32 \pm 15 \text{ MeV}, \\
f_{B_s} &= 207(\phantom{0}9) \pm 10 \pm 32 \pm 22 \text{ MeV}, \\
f_D &= 208(\phantom{0}9) \pm 11 \pm 33 \pm 12 \text{ MeV}, \\
f_{D_s} &= 230(\phantom{0}8) \pm 10 \pm 28 \pm 18 \text{ MeV}.
\end{aligned} \qquad (4.3)$$

The uncertainty in parentheses is statistical; the others are systematic. From left to right, they are due to the following sources:

---
[*]Because of helicity mismatch, the rate is proportional to $m_l^2$, which makes the measurement difficult. This example is worth pursuing—at least pedagogically—because it is so simple.

1. Fitting, interpolation, and extrapolation. The $t$ dependence of the numerical two-point functions is fit to eqs. (2.2) and (2.4) once the lowest-lying pseudoscalar has been isolated. The mass of the heavy quark is adjusted by interpolating to $m_b$ or $m_c$; the mass of the light quark is adjusted by extrapolating to $(m_u+m_d)/2$ or $m_s$. These could be reduced somewhat in concert with a reduction in the statistical error.

2. Large $m_b$ effects. Two $1/m_b$ contributions modify the static limit, the kinetic energy and a chromomagnetic $i\boldsymbol{\Sigma}\cdot\boldsymbol{B}$ term. Owing to lattice artifacts in the standard lattice action, the quark mass is tuned so that the kinetic energy has the correct strength, the chromomagnetic term is too weak.[17, 19] This could be reduced with an improved action, as done in Ref. 20. The results with the improved action agree extremely well with eq. (4.3), especially when one compares $f_P/f_\pi$ from both.[18, 20]

3. Uncertainty in the conversion from lattice units to MeV. As in Ref. 5, it turns out that $(af_\pi/am_\rho)_{\text{Ref. 18}} \neq (f_\pi/m_\rho)_{\text{expt}}$. This could be an artifact either of non-zero lattice spacing or of finite volume, but these possibilities are unlikely because Ref. 18 agrees with Ref. 5, which extrapolates these two effects away. Another culprit could be the quenched approximation, which is, perhaps, more likely.

A remarkable feature of Ref. 18 is that the number of systematic uncertainties quoted equals the number of authors.

In ratios many of the errors cancel, because of statistical and systematic correlations. The result from Ref. 18

$$\frac{f_D}{f_{D_s}} = \frac{f_B}{f_{B_s}} = \frac{f_B}{f_D} = \frac{f_{B_s}}{f_{D_s}} = 0.90 \pm 5\%. \qquad (4.4)$$

is easy to remember. If $f_{D_s}$ were experimentally determined to 5%, eq. (4.4) would perhaps be more relevant than eq. (4.3).

The uncertainty estimates do not explicitly include quenched, finite-volume, or non-zero lattice spacing errors. As indicated above, however, some of these errors are implicitly included in the estimates quoted. From the studies of the static limit[13, 14] one expects the volume dependence to be insignificant once the volume is "large enough." The lattice-spacing dependence, on the other hand, is surprisingly large.

The results shown in eq. (4.3) may disagree with previous lattice calculations. Some older results were higher, quoting values larger than 300 MeV for $f_B$. Such numbers came typically from early calculations in the static limit, neglecting the dependence on the heavy quark mass. In addition, the early studies were at larger lattice spacings and often used operators that were unsuccessful in isolating the lowest-lying states. Other older results were lower. These results typically started with heavy quark that were relatively light, and extrapolated. These extrapolations were done using an incorrect normalization of the current. The correct normalization is now understood[17, 19] and Ref. 18, for example, uses it. The difference is most noticeable on coarse lattices; the impact of the correct normalization and an associated mass shift[17, 19] is shown in Fig. 5, using numerical data from Ref. 21.

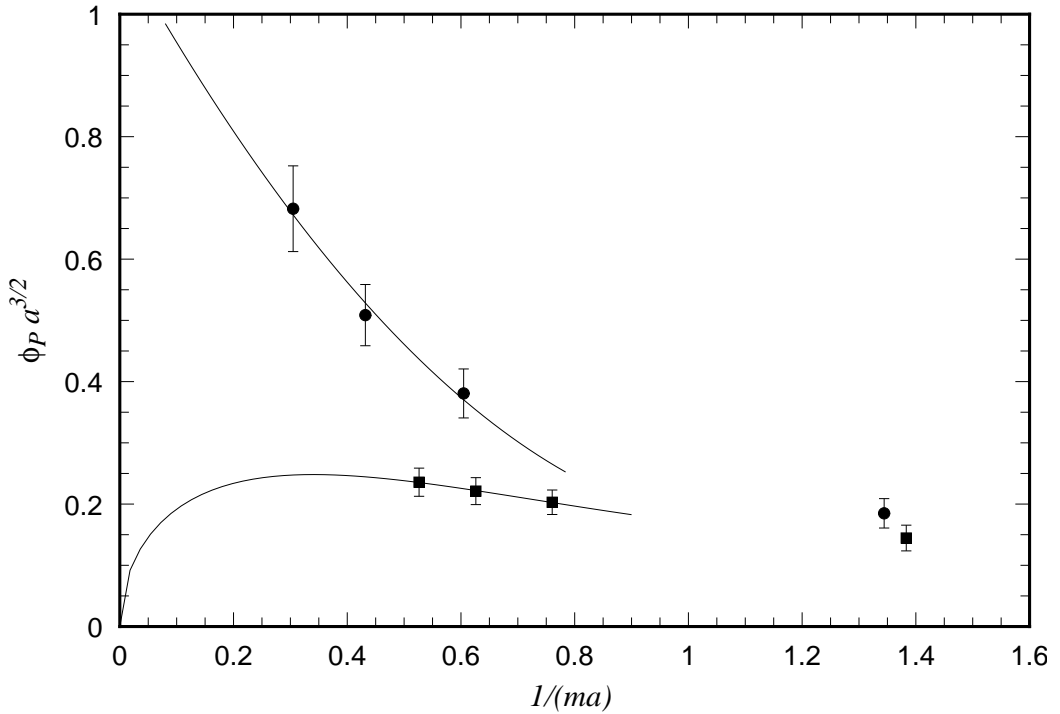

Figure 5: Putative calculations of $\phi_P = \sqrt{m_P} f_P$, where $P$ denotes a heavy-light pseudoscalar meson, as a function of (inverse) mass. The squares denote an incorrect current normalization, which systematically underestimates $\phi_P$. The circles use a current normalization and mass definition derived in Ref. 17. The curves indicate the large mass behavior in each case.

### 4.2 Semileptonic Decays: $A_1^{B \to D^*}(q^2)|V_{cb}|$ and $A_1^{B \to \rho}(q^2)|V_{ub}|$

The rates of semileptonic decays exceed those of pure leptonic decays, because they do not suffer from helicity suppression. They therefore lend themselves particularly well to the determination of elements of the CKM matrix. The rates are measurable and the reliability of theoretical calculations is better than for nonleptonic decays (sect. 4.4). For example, the best determination of $|V_{us}|$ comes from $K \to \pi l \nu$, and the best determination of $|V_{cb}|$ comes from $B \to D^* l \nu$.

We shall focus on mesons, because they are easier than baryons to study, both experimentally and theoretically. A generic semileptonic decay can be denoted $A \to X l \nu$, where $A$ is a flavored meson. The process is depicted in Fig. 6. The differential decay rate follows the pattern of eq. (1.1):

$$\frac{d\Gamma}{dq^2} = \left[ \frac{G_F^2 \lambda^{3/2}}{192 \pi^3 m_A^3} \right] |f_+(q^2)|^2 |V_{ax}|^2, \quad (4.5)$$

when $X$ is a pseudoscalar meson, and

$$\frac{d\Gamma}{dq^2} = \left[\frac{G_F^2 \lambda^{1/2} q^2}{64\pi^3 m_A}\right] |A_1(q^2)|^2 |V_{ax}|^2, \quad (4.6)$$

when $X$ is a vector meson. In eqs. (4.5) and (4.6), $q^2$ is the invariant mass of the virtual $W$ [$0 < q^2 \leq q^2_{\max} = (m_A - m_X)^2$], $V_{ax}$ is the element of the CKM matrix associated with the quark-$W$ vertex in Fig. 6, and $\lambda = (m_A^2 + m_X^2 - q^2)^2 - 4m_A^2 m_X^2$. For brevity and a reason explained below, eq. (4.6) is valid only for $q^2$ near $q^2_{\max}$. The form factors $f_+$ and $A_1$ are defined by hadronic matrix element of the $V - A$ current

$$J_\mu = \bar{x}\gamma_\mu(1 - \gamma_5)a \quad (4.7)$$

turning flavor $a$ into flavor $x$. When $X$ is a pseudoscalar meson

$$\langle X|J_\mu|A\rangle = f_+(q^2)(p + p')_\mu + f_-(q^2)(p - p')_\mu, \quad (4.8)$$

where $p$ ($p'$) is the initial (final) state meson's momentum and $q = p - p' = p_l + p_\nu$. Similarly, when $X$ is a vector meson there are four independent form factors:

$$\langle X|J_\mu|A\rangle = \epsilon_\lambda^* \left[\frac{2\varepsilon_{\mu\lambda\rho\sigma}p_\rho p'_\sigma}{m_A + m_X}V(q^2) - \delta_{\mu\lambda}(m_A + m_X)A_1(q^2)\right.$$
$$\left. + \frac{(p + p')_\mu p_\lambda}{m_A + m_X}A_2(q^2) - \frac{2m_X(p - p')_\mu p_\lambda}{q^2}A(q^2)\right], \quad (4.9)$$

where $\epsilon_\lambda^*$ is the polarization vector of the final-state meson. The form factors $f_-$ and $A$ do not appear in the expressions for the differential decay rates because the lepton mass has been neglected; $A_2$ and $V$ do not appear for $q^2$ near $q^2_{\max}$ because they are suppressed by a higher power of $\lambda$.

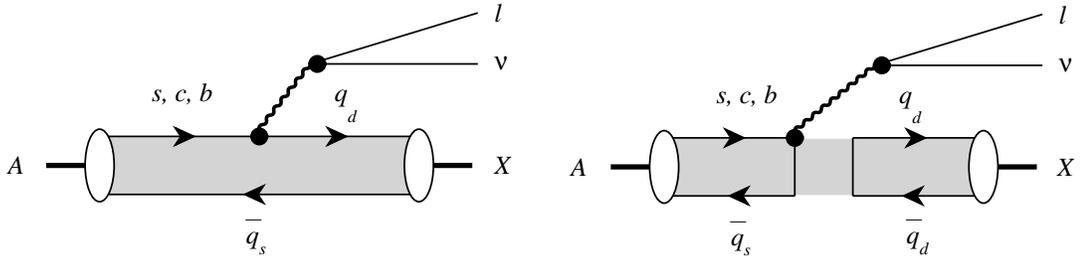

Figure 6: Quark-flow diagrams for meson semileptonic decays. For the weak interactions, the diagram may be interpreted as a Feynman diagram. The strong interactions binding quarks into mesons must be treated nonperturbatively, however, as indicated by the gray shading. The second diagram contributes only when $X$ is an isoscalar. It is usually neglected, because it is difficult to calculate and because diagrams similar to Fig. 2(d) are omitted in the quenched approximation anyway.

Table 1: Semi-leptonic decays and the CKM matrix elements they determine. For brevity only pseudoscalar final states are listed; vector final states are $\rho$, $K^*$ and $D^*$, as appropriate.

| $A \to X$ | $V_{ax}$ | COMMENT |
|---|---|---|
| $K \to \pi$ | $V_{us}$ | calibrate quenched approximation |
| $D \to \pi$ | $V_{cd}$ | uncertainty in $|V_{cd}|$ dominated first by BR($D \to \pi l \nu$), then by $f_+$ |
| $D \to K$ | $V_{cs}$ | uncertainty in $|V_{cs}|$ dominated by $f_+$ |
| $B \to D$ | $V_{cb}$ | test/compute corrections to heavy quark limit |
| $B \to \pi$ | $V_{ub}$ | $\rho$ final state more useful; cf. text |

Table 1 lists a variety if semileptonic decays and their utility in either testing numerical lattice QCD methods or extracting CKM matrix elements. For $B$ decays two entries are of note, depending on whether the quark-level decay is $b \to c$ or $b \to u$.

For $B \to D^{(*)}$ both the charm and bottom quarks are reasonably heavy and one can apply heavy-quark symmetry. The kinematic endpoint $q^2_{\max} = (m_B - m_{D^{(*)}})^2$ is especially interesting, because then one can determine the $B \to D^* l\nu$ differential decay rate up to corrections of order[22, 23] $1/m^2_{D^*}$. A similar analysis shows that the leading correction to the $B \to Dl\nu$ differential decay rate is O($1/m_D$). For $q^2 < q^2_{\max}$ the corrections are O($1/m_{D^{(*)}}$) for both final states. Using estimates from QCD sum rules for the $1/m^2_{D^*}$ and $1/m^2_B$ corrections to $A_1(q^2_{\max})$ enables one to limit the theoretical uncertainty on $|V_{cb}|$ to 4%. It seems unlikely that lattice QCD can improve on this bottom line any time soon, although verification of the QCD sum rule calculations would be important. Another contribution that lattice QCD can make is a model-independent determination of the $q^2$ dependence. This would assist the extrapolation of the experimental data towards the statistics-poor endpoint, possibly reducing the overall uncertainty on $|V_{cb}|$. Exploratory results in this direction have appeared recently.[24, 25]

Lattice QCD can make a more significant impact on the determination of $V_{ub}$. Since the $\pi$ or $\rho$ is light, heavy-quark symmetry could only be used to relate, say, $D \to (\pi$ or $\rho)$ form factors to $B \to (\pi$ or $\rho)$ form factors.[26] As above either models or lattice QCD would be needed to compute the $1/m_D - 1/m_B$ corrections. Strictly speaking, the end result would be $|V_{ub}/V_{cd}|$. It seems more reasonable to use lattice QCD to calculate the form factors and use the experiments to determine $|V_{ub}|$ and $|V_{cd}|$ separately. As pointed out in Ref. 27 the cleanest procedure is to use $\rho$ final states with $q^2$ near $q^2_{\max}$. The calculations are most reliable at $q^2_{\max}$, because $q^2 < q^2_{\max}$ is obtained for $\boldsymbol{p}' \neq \boldsymbol{0}$, and when $|\boldsymbol{p}'|a \sim 1$ there are additional lattice artifacts. Near $q^2_{\max}$ the phase spaces suppression is less drastic for vector mesons than for pseudoscalar mesons. No calculation of $A_1^{B \to \rho}(q^2)$ with a thorough error analysis is available yet, although it seems to feasible to complete a calculation with 5–10% errors by the time experimental data for $d\Gamma/dq^2$ become available.

Let us sketch how this will come about, starting from the estimates of the systematic uncertainties for $D \to K^{(*)}l\nu$ in Refs. 28 and 27. One ought to be able to reduce the 10–20% statistical uncertainty of published calculations[29, 28, 30, 27] to 2-5%. At that level it is possible to treat the systematic quantitatively. (Refs. 28 and 27 made semi-quantitative estimates; other papers[29, 30] felt that their large statistical uncertainties made estimates of systematic errors premature.) The previous 20–40% uncertainty from O($a$) effects should be reduced to below the statistical error, by extrapolating in $a$. The 5–20% uncertainty owing to inadequate knowledge of quark masses should fall to the level limited by mass calculations, which is presently estimated to be 2–6%.[4] Finally, although volume dependence is probably not a problem, momentum and, hence, $q^2$ take on discrete values in a finite volume. A variety of volumes would make available more values of $q$;

### 4.3 $B_q^0$-$\bar{B}_q^0$ Mixing: $f_B^2 B_B |V_{td}|^2$ and $f_{B_s}^2 B_{B_s} |V_{ts}|^2$

Neutral-meson mixing is interesting from the point of view of the CKM matrix, because it offers a handle on the third row. The rate of mixing is related to

$$x_d = \frac{\Delta m_{B^0}}{\Gamma_{B^0}} = \left[\frac{G_F^2 m_t^2 \tau_B}{16\pi^2 m_B} f_2(m_t^2/m_W^2)\right] \eta_{\text{pQCD}} \tfrac{8}{3} m_B^2 f_B^2 B_B |V_{td}^* V_{tb}|^2, \quad (4.10)$$

where

$$\tfrac{8}{3} m_B^2 f_B^2 B_B = \langle \bar{B}^0 | \bar{b}_i \gamma_\mu (1-\gamma_5) d_i \bar{b}_j \gamma_\mu (1-\gamma_5) d_j | B^0 \rangle. \quad (4.11)$$

Similar expressions hold for the $B_s$ meson, substituting an $s$ quark for the $d$ quark throughout. The perturbative QCD factor $\eta_{\text{pQCD}}$ has been grouped outside of the bracket of known factors, even though it is known, because both $\eta_{\text{pQCD}}$ and $B_B$ depend on the renormalization scheme, but the product $\eta_{\text{pQCD}} B_B$ does not. Even though the top-quark mass $m_t$ is not yet known, the dependence on it is grouped with the known factors, because it should be known soon; the function $f_2$ is known.

The peculiar but traditional notation $B_B$ is useful for lattice QCD calculations, because $B_B$ is then a ratio of matrix elements for which many uncertainties cancel. Although the analogous quantity in the kaon system represents one of the most reliable lattice QCD calculations,[31] calculations of $B_B$ are still exploratory.[32] At the 20–40% level, there is no evidence yet for a significant deviation from the naive expectation $B_B = 1$.

The dependence on the top-quark mass and some other "known" factors, cancel in the ratio, leaving

$$\frac{x_d}{x_s} = \left[\frac{\tau_B m_B}{\tau_{B_s} m_{B_s}}\right] \frac{f_B^2 B_B}{f_{B_s}^2 B_{B_s}} \left|\frac{V_{td}}{V_{ts}}\right|^2. \quad (4.12)$$

Hence, an experimental measurement of $x_d/x_s$ together with a lattice QCD calculation of $f_B^2 B_B/(f_{B_s}^2 B_{B_s})$ determines $|V_{td}/V_{ts}|$. As in eq. (4.4) the uncertainty in the $B$-to-$B_s$ ratio should be smaller than in numerator or denominator separately.[32]

### 4.4 Nonleptonic Decays

Nonleptonic decays, such as $B \to J/\psi K_S$ or $B \to \pi^+\pi^-$, receive almost all of the attention in discussions of $CP$ violation. A serious obstacle to the treatment of nonleptonic decays is the presence of two (or more) hadrons in the final state. The technical aspect is the difficulty of separating the particles in the finite volume. The conceptual aspect is the determination of final-state phase shifts from purely real quantities computed in Euclidean field theories.[33, 34] It is rigorously known[35] how to determine the resonance properties of the $\rho$, which decays through an interaction in the QCD Hamiltonian. The stumbling block for weak $B$ decays is evidently the application of the ideas in Ref. 35 when the particle decays through an interaction being treated as a perturbation. Note that these difficulties do *not* stem from the lattice cutoff, but from other features, finite volume and imaginary time, introduced to make the computational method tractable. Nevertheless, until these issues are resolved, lattice results for nonleptonic decays probably will not warrant attention from non-experts.

With the lattice QCD calculations discussed above, however, it *will* be able to determine the angles of the unitarity triangle, as discussed in sect. 5

### 4.5 Qualitative Information

An interesting qualitative result for the $B$ meson is its valence wave function. The intriguing result[36] is that the wave functions in the static limit are completely consistent with wave functions of the semi-relativistic potential model with Hamiltonian

$$H = \sqrt{p^2 + m^2} + V_{q\bar{q}}(x), \qquad (4.13)$$

where $m$ is the (reduced) mass of the light quark and $V_{q\bar{q}}(x)$ is Buchmüller-Tye potential or any other empirical potential consistent with asymptotic freedom, linear confinement, and quarkonium phenomenology. Because of the relativistic kinetic energy, the wave functions are much broader than in a nonrelativistic model. In particular, the true wave function seems to be much broader than those used in phenomenological quark models.

## 5 FUTURE PROSPECTS

The standard model has around 20 parameters and, in the long run, precision lattice QCD calculations are needed to determine half of them ever more precisely.[3] In particular, properties of the $B$ meson are needed to pin down the four parameters associated with the CKM matrix. Indeed, in the standard 3-generation parameterization $|V_{us}|$, $|V_{cb}|$, and $|V_{ub}|$ yield (to good approximation) $\theta_{12}$, $\theta_{23}$, and $\theta_{13}$, respectively. These three together with $|V_{td}|$ yield the phase $\delta$ responsible for $CP$ violation. Hence, semileptonic decays and mixing of the $B$ meson, together with the calculations described above, are essential to determining three out of the four CKM parameters.

To put an even finer point on this observation, consider the unitarity triangle. The magnitudes of its sides are $|V_{ud}V_{ub}^*|$, $|V_{cd}V_{cb}^*|$, and $|V_{td}V_{tb}^*|$. Sect. 4.2 shows how to determine $|V_{cb}|$ and $|V_{ub}|$ with semileptonic decays; a similar technique for charm decays

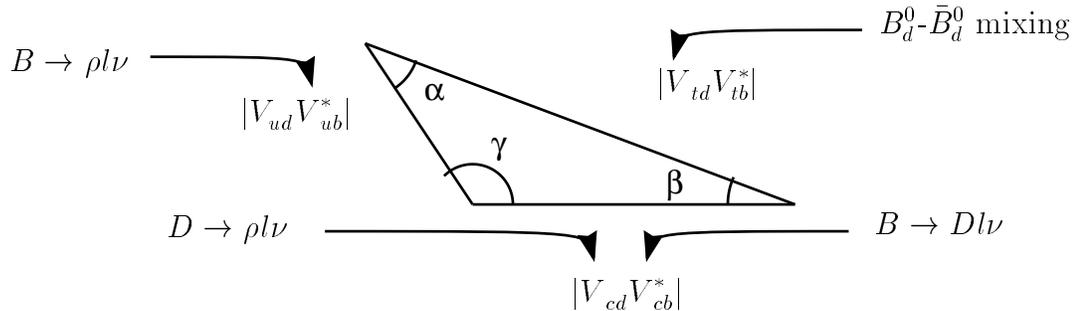

Figure 7: The unitarity triangle and the $B$ properties needed to determine the sides. $|V_{ud}|$ is known from $K \to \pi l \nu$ and $|V_{tb}|$ is known from three-generation unitarity.

determines $|V_{cd}|$. Sect. 4.3 shows how to determine $|V_{td}|$ from neutral-meson mixing. Now assume three-generation unitarity. (Eq. (4.10) already does so.) That implies that the three sides form a triangle, as shown in Fig. 7. It also implies $|V_{tb}| = 1$, $|V_{ud}| = 0.976$, and $|V_{ts}| = |V_{cb}|$; the latter can improve the determination of $|V_{td}|$ through $|V_{td}/V_{ts}|$, cf. eq. (4.12). Of all these CKM matrix elements $|V_{ub}|$ is the most poorly known, but the experimental and theoretical work of the next few years will improve the determination. Once is it precise enough, all three sides will be known, and, as any child will tell you, then the angles are known too.

Most theoretical descriptions of $CP$ asymmetries cast them as measurements of the angles $\alpha$, $\beta$, and $\gamma$. But three-generation unitarity is often assumed and penguin contributions are almost always assumed to be unimportant. Using the calculations discussed above, however, one need only assume three-generation unitarity to determine $\alpha$, $\beta$, and $\gamma$. Because the measurements involved all conserve $CP$, they will most likely be available before the $CP$ asymmetries are. If that is indeed so, it is more accurate to say that measurements of $CP$ asymmetries test the CKM theory of $CP$ violation, than to say that they determine the CKM parameters.

## ACKNOWLEDGEMENTS

I would like to thank M. Uchima for nursing my ailing Exabyte tapes, while I was enjoying the mountain air of Snowmass.